\documentclass[a4paper]{article}
\pdfoutput=1
\usepackage{INTERSPEECH2021}
\usepackage{multirow}
\usepackage{xcolor, tabu, cite}

\title{Teacher-Student MixIT for Unsupervised and Semi-supervised Speech Separation}
\name{Jisi Zhang$^1$, C\u{a}t\u{a}lin Zoril\u{a}$^2$, Rama Doddipatla$^2$ and Jon Barker$^1$}
\address{
  $^1$University of Sheffield, Department of Computer Science, Sheffield, UK \\
  $^2$Toshiba Cambridge Research Laboratory, Cambridge, UK}
    \email{\{jzhang132,j.p.barker\}@sheffield.ac.uk, \{catalin.zorila,rama.doddipatla\}@crl.toshiba.co.uk}

\begin{document}

\maketitle
\begin{abstract}
In this paper, we introduce a novel semi-supervised learning framework for end-to-end speech separation. The proposed method first uses mixtures of unseparated sources and the mixture invariant training (MixIT) criterion to train a teacher model. The teacher model then estimates separated sources that are used to train a student model with standard permutation invariant training (PIT). The student model can be fine-tuned with supervised data, i.e., paired artificial mixtures and clean speech sources, and further improved via model distillation. Experiments with single and multi channel mixtures show that the teacher-student training resolves the over-separation problem observed in the original MixIT method. Further, the semi-supervised performance is comparable to a fully-supervised separation system trained using ten times the amount of supervised data.

\end{abstract}

\section{Introduction}

Speech separation, the process of segregating an individual target speaker from an overlapping interferer, has made great progress, driven by the power of deep learning.
As a front-end process, it has benefited many applications, including speaker diarization, speaker verification and multi-talker speech recognition.
The performance of speech separation has been studied in various simulated acoustic conditions including close-talking and distant microphone conditions.
Initially, for single-channel separation, evaluations focused on synthetic anechoic mixture data produced by simply weighting and summing two non-overlapped speech signals~\cite{hershey2016deep,isik2016single,luo2019conv}.
More recently, attention has turned to simulated noisy and reverberant mixture data for developing far-field multi-channel speech separation approaches~\cite{Wichern_2019,Maciejewski_2020,drude2019smswsj}.
Specifically, speech and pre-recorded noise are mixed using room impulse responses (RIRs) simulated using room acoustic simulation toolkits~\cite{allen1979image,scheibler2018pyroomacoustics}.

Although separation models trained with simulated data have achieved impressive performances when evaluated with matched simulated data, they often perform poorly on real mixtures due to mismatches between the real and simulated environments. This cannot be fixed simply by retraining with `real data' as in real scenarios there is generally no access to isolated ground truth signals.
Recently, a novel \textit{fully-unsupervised} end-to-end separation technique, known as mixture invariant training (MixIT), has been proposed as a solution to this problem~\cite{wisdom2020unsupervised}. MixIT uses a mixture of mixtures during training (i.e., artificial mixtures of the real scenes that are to be separated). A training strategy allows it to infer individual sources within each mixture. However, the technique can suffer from an `over-separation' problem caused by a training data mismatch: the training data - a mixture of mixtures - will contain more sources than the individual mixtures that it is being trained to segregate.

Teacher-student approach has also been studied to address the mismatch between the real and simulated environments recently~\cite{tzinis2019unsupervised,drude2019unsupervised,bando2019deep,lam2020mixup}.
When multiple microphones are available, a standard approach is to use an unsupervised spatial clustering approach as a teacher for a single-channel separation network student~\cite{tzinis2019unsupervised,drude2019unsupervised,bando2019deep}.
However, this approach relies on multimodal training data and has not been evaluated with recent developed end-to-end separation models.
Mixup-Breakdown training combines teacher-student learning and data augmentation to adapt separation models to mismatched interference~\cite{lam2020mixup}. But the teacher model is still trained with simulated supervised data.


This paper presents a novel semi-supervised learning framework for the end-to-end speech separation task. 
To address the over-separation issue, we first propose an unsupervised algorithm that combines the teacher-student learning and MixIT criterion, which we call TS-MixIT.
Specifically, the algorithm uses mixture of mixtures to train a teacher model with the MixIT criterion. Then the teacher model processes original mixtures to generate separated sources, from which speech signals are selected with an energy restriction. The selected signals are used to train a student model with a smaller number of output channels with the PIT criterion~\cite{kolbaek2017multitalker}. Second, we investigate how to further improve the unsupervised trained model in situations where a small amount of ground-truth separated training data is available. Inspired by previous work demonstrating that pre-trained unsupervised models can be fine-tuned for downstream tasks~\cite{devlin2018bert,oord2018representation,Schneider2019}, we fine-tune the unsupervised student model to be more task-specific. Finally, following recent work on model distillation \cite{chen2020big}, we use a second teacher-student training stage to generate a more compact task-specific model, increasing performance and efficiency in the process.

The rest of paper is organised as follows. In Section~\ref{sec:method}, we introduce the proposed teacher-student MixIT training method. Section~\ref{sec:implement} presents single- and multi-channel separation experiments in anechoic and noisy reverberant conditions respectively. Results and analysis are presented in Section~\ref{sec:result}. Finally, this paper is concluded in Section~\ref{sec:conclude}.

\vspace{-5pt}
\section{Methods}
\label{sec:method}
End-to-end speech separation models take waveforms as both input and output and have made great progress in both single-channel and multi-channel cases recently~\cite{luo2019conv,luo2020dual,zeghidour2020wavesplit,nachmani2020voice,luo2020end,zhang2021time}.
The architecture usually consists of an encoder, a decoder and a separator module for estimating a mask for each source.
Permutation invariant training~\cite{kolbaek2017multitalker} has become a standard training framework for training a separation network in a supervised way.
Given a mixture $x$ containing up to $C$ sources and its corresponding ground truth signals $\mathbf{s}$, the separation model takes the mixture $x$ as input and predicts $M=C$ sources.
Since the order of the predicted signals is arbitrary, $C!$ permutations of prediction to ground truth pairs exist for computing the signal reconstruction loss. 
PIT addresses this problem by using the permutation which gives the smallest loss computed over the entire utterance. The smallest loss will be used to back-propagate to optimise the model parameters.

\subsection{Mixture Invariant Training}
\begin{figure}[htp]
    \centering
    \includegraphics[width=0.47\textwidth]{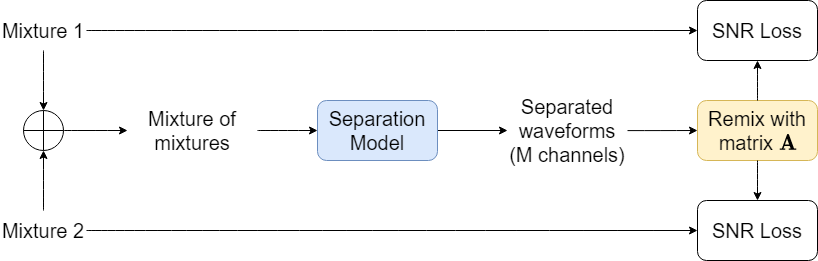}
    \caption{MixIT training algorithm}
    \label{fig:mixit}
\end{figure}

The mixture invariant training criterion, as illustrated in Figure~\ref{fig:mixit}, is designed for training separation model in an unsupervised fashion~\cite{wisdom2020unsupervised}.
The MixIT draws two mixtures at random without replacement from an unsupervised dataset. The two selected mixtures form a mixture of mixtures by adding them together: $\overline{x}=x_1+x_2$.
An end-to-end separation model takes $\overline{x}$ as input and predicts $M\geq2C$ source signals, which will be used to reconstruct the two mixtures.
The unsupervised MixIT loss is computed between the estimated sources $\hat{\mathbf{s}}$ and the input mixtures $x_1,x_2$ as follows:
\vspace{-5pt}
\begin{equation}
    \mathcal{L}_{\mathrm{MixIT}}(x_1,x_2,\hat{\mathbf{s}})=\min_{\mathbf{A}}\sum_{i=1}^2\mathcal{L}(x_i,[\mathbf{A}\hat{\mathbf{s}}]_i),
\end{equation}
where $\mathbf{A} \in \mathbb{B}^{2 \times M}$ is a mixing matrix whose elements along each column sum to 1 assigning each $\hat{s}$ to either $x_1$ or $x_2$.
The mixing matrix combines the estimated sources after the decoder in the separation model.

The signal-level loss function between a reference $y \in \mathbb{R}^{T}$ and estimate $\hat{y} \in \mathbb{R}^{T}$ from the separation model is the negative threshold SNR:
\vspace{-5pt}
\begin{equation}
\begin{split}
        \mathcal{L}(y, \hat{y}) & = -10\mathrm{log}_{10} \frac{\|y\|^2}{\|y-\hat{y}\|^2+\tau\|y\|^2} \\ 
        &= 10\mathrm{log}_{10}(\|y-\hat{y}\|^2+\tau\|y\|^2) - 10\mathrm{log}_{10}\|y\|^2,
\end{split}
\end{equation}
where $\tau=10^{-\mathrm{SNR}_{max}/10}$ acts as a soft threshold that clamps the loss at $\mathrm{SNR}_{max}\mathrm{=30~dB}$, which is the same value as in~\cite{wisdom2020unsupervised}. $T$ is the signal's length in samples.
\vspace{-5pt}
As described in the original paper, a mixture consistency constraint~\cite{wisdom2019differentiable} has been introduced during training the MixIT model, which forces the network to predict sources that sum up to the mixture. This constraint solves the following optimization problem to find mixture consistency separated sources $\hat{s}_m$ given initial separated sources $\underline{s}_m$:
\begin{equation}
        \min_{\hat{s} \in \mathbb{R}^{M \times T}} \frac{1}{2} \sum_m \|\hat{s}_m - \underline{s}_m \|^2 \  \text{subject} \ \text{to} \sum_m\hat{s}_m = x.
\end{equation}
The closed-from solution of this problem is:
\begin{equation}
    \hat{s}_m = \underline{s}_m + \frac{1}{M} (x - \sum_{m^\mathrm{'}} \underline{s}_{m^\mathrm{'}}).
\end{equation}

Although the MixIT framework was shown to perform well,
 it suffers from mismatch and over-separation issues.
Firstly, since the number of speakers during training is always larger than that during inference, there is a severe mismatch between training and evaluation.
Next, the over-separation issue will appear when the number of output channels from the separation model is larger than the number of speakers in a mixture sample, which is common in a MixIT trained network. This also causes problem to select speech signals from output channels.

\subsection{Teacher-student training algorithm}
\begin{figure}[htp]
    \centering
    \includegraphics[width=0.47\textwidth]{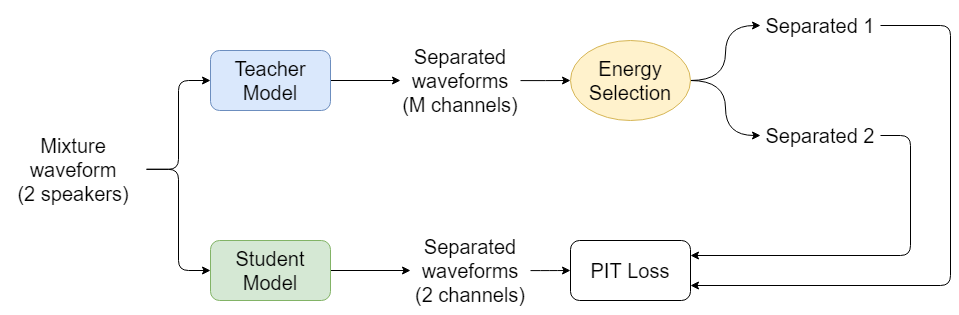}
    \caption{Teacher-student MixIT training algorithm}
    \label{fig:ts_mixit}
\end{figure}

In order to solve the mismatch and over-separation problems, we wish to reduce the number of output channels to match the number of sources in the mixture signals.
Based on this, we propose an unsupervised training algorithm that combines both MixIT and PIT criterion.
We first train a teacher model on mixture of mixtures using the MixIT criterion.
Then, the teacher model is used to process the original mixtures to generate M isolated signals, of which speech signals are selected with an energy restriction.
We assume that the number of speakers in the mixture is known and the same number of output streams with the highest energies will be selected as the pseudo-targets.
Then, the student model is trained to separate the same mixture sample into the selected separated sources with the PIT loss to minimize the signal reconstruction loss.
Both the teacher model and the student model have a similar architecture except the layer for estimating masks.
The teacher-student training process for mixtures of two speakers is illustrated in Figure~\ref{fig:ts_mixit}.

In the student model, the application of the mixture consistency constraint depends on the noise condition. It is kept when the separation is under the anechoic condition since the separated sources should sum up to the original mixture. However, in the noisy and reverberant condition, it is removed since the student model only outputs speech signals and discards noise signals.
\subsection{Fine-tuning}
The previous teacher-student learning can be considered as an unsupervised pre-training process. Unsupervised pre-training teaches the model to learn a general representation that can be transferred to downstream tasks~\cite{devlin2018bert,oord2018representation,Schneider2019}. The unsupervised pre-training can also alleviate the overfitting problem of supervised learning when the amount of labeled data is small.

When a small amount of parallel mixture and target signals is available, the entire pre-trained student model can be fine-tuned without replacing or discarding any layers. In this work, this simple strategy is used to show that the model can benefit from the unsupervised pre-training.
There are more complex strategies for model fine-tuning or transfer learning in the literature, e.g., freezing or replacing layers, and adjusting the learning rate~\cite{tzinis2020two,zhang2020learning}, which can be explored in our future work.

\subsection{Knowledge distillation via unlabeled examples}
It has been shown that deeper and bigger models can benefit more from the task-agnostic use of unlabeled data, while a task-specific model may be best structured in a different way and have a smaller model size~\cite{chen2020big}. With this observation as motivation, to further improve the network for the separation task, we use the fine-tuned model as a teacher to generate pseudo-targets again for training a student network with a different structure. 
A more compact model previously shown to be better designed for the separation task~\cite{luo2020dual} is selected as the student model. The teacher network is fixed during the distillation and only the parameters of the student network are optimised.

\section{Experiment Setup}
\label{sec:implement}
\subsection{Data simulation}
The proposed training framework is evaluated on different tasks under various acoustic conditions, using both single and multi-channel data. The single-channel separation task is evaluated under anechoic conditions, and the multi-channel separation task is evaluated under noise and reverberation.
In the latter case, the separation model is trained to separate reverberant sources from the noisy mixture.

For the single-channel experiments we use the anechoic wsj0-2mix dataset~\cite{hershey2016deep}.
For the multi-channel case, the evaluation is performed on the WHAMR! dataset with a stereo (2-channels) configuration, which consists of simulated noisy and reverberant 2-speaker mixtures~\cite{Maciejewski_2020}.
WHAMR! is based on Wall Street Journal (WSJ) data, which is mixed with noise recorded in various urban environments~\cite{Wichern_2019}, and reverbed using artificial room impulse responses~\cite{scheibler2018pyroomacoustics}.
All data have 8 kHz sampling rate.
For both the single-channel and the multi-channel case, there are 20k sentences from 101 speakers for training, and 3k sentences from 18 speakers for testing.
The dynamic remixing strategy~\cite{zeghidour2020wavesplit} is used to train the MixIT model, and the first 2k sentences without dynamic remixing are used for the fine-tuning task.
The speakers in the test set do not appear during training of the speech separation system.

The MixIT model is trained with two strategies for creating the mixture of mixtures: i) Both mixtures always consist of two speakers (\emph{2-src}) such that the mixture of mixtures always contain four sources. ii) One of the mixtures use either a single or two speakers (\emph{1or2-src}) such that the mixture of mixtures contain three or four sources. In the later case, 10\% of training data is simulated with a single speaker.

\subsection{Separation network configurations}
For the single-channel separation task, the Conv-TasNet~\cite{luo2019conv} architecture is employed to train the TS-MixIT model.
Conv-TasNet's hyper-parameters are set as those that produced best performance in the original paper namely, $N=256$, $B=128$, $H=256$, $L=20$, $R=4$, $X=7$. The number of output channels $M$ for the teacher model and the student model in the anechoic condition is set as 4 and 2, respectively.
For the multi-channel separation task, the end-to-end multi-channel separation network proposed in~\cite{Zhang2020end} is employed, and the hyper-parameters are as follows: $N=256$, $S=36$, $R=3$, $X=7$, $L=16$. The number of output channels $M$ for the teacher model and the student model in the noisy and reverberant condition is set as 8 and 2, respectively.
Each utterance is split into multiple segments with a fixed length of 4 seconds.
All models are trained with the Adam optimiser~\cite{kingma2014adam} with a learning rate of 1e-3 and a batch size of 8.

For knowledge distillation, we used separation models outperforming the Conv-TasNet in the supervised learning case, while maintaining a smaller model size. The dual-path RNN (DPRNN)~\cite{luo2020dual} and the U-Convolutional block (U-ConvBlock) based multi-channel separation networks~\cite{zhang2021time} have been selected for the single- and the multi-channel tasks, respectively.


\subsection{Acoustic model}
To evaluate the speech recognition performance, two acoustic models (AMs) have been trained using the Kaldi speech recognition toolkit~\cite{povey2011kaldi}. One model (AM1) was trained on roughly 80~hrs of clean WSJ-SI284 data plus the WHAMR! single-speaker noisy reverberant speech, and the other one (AM2) was trained on the data used for AM1 plus the separated signals from the training mixtures of WHAMR! processed by the proposed multi-channel model with fine-tuning and knowledge distillation.
The audio is downsampled to 8 kHz to match the sampling rate of the data used for the separation experiments.
The acoustic model topology is a 12-layer factorised TDNN~\cite{povey2018semi}, where each layer has 1024 units, and it is trained using 40-dimensional MFCCs and 100-dimensional i-Vectors.
Recognition is performed using a 3-gram language model provided in the baseline Kaldi WSJ recipe.
With our set-up, the word error rate (WER) results obtained with AM1 on the standard clean WSJ Dev93 and Eval92 are 7.2\% and 5.0\%, respectively.

\section{Results and Analysis}
\label{sec:result}

\subsection{Single-channel separation}

\begin{table}[!b]
\vspace*{-3mm}
\centering
\caption{Results for anechoic single-channel separation using unsupervised learning (Conv-TasNet architecture)}
\label{tab:ts-mixit_anechoic}
\begin{tabu}{lccc}
\hline
\textbf{System}     & \textbf{Mixes}       & \textbf{M} & \textbf{SI-SNRi (dB)} \\ \hline
{MixIT (Energy)}    & 2-src                & 4          & 9.0                   \\
{MixIT (Oracle)}    & 2-src                & 4          & 9.8                   \\
{TS-MixIT (Proposed)}          & 2-src                & 2          & 10.4                   \\ \hline
{MixIT (Energy)}    & 1or2-src             & 4          & 12.0                  \\
{TS-MixIT (Proposed)}          & 1or2-src             & 2          & 12.6                  \\ \hline
\rowfont{\color{gray}}
{Supervised (10\%)} & -                & 2          &   11.0                 \\ \hline
\end{tabu}
\end{table}
Table~\ref{tab:ts-mixit_anechoic} shows the single-channel separation performance of the proposed method (TS-MixIT) in anechoic conditions. 
The scale-invariant signal-to-noise ratio (SI-SNR)~\cite{le2019sdr} is used to measure the separation performance. The baseline MixIT model is implemented with the Conv-TasNet model and two separated signals with the highest signal energies are selected from the $M$ output channels during inference. We also conducted an oracle experiment by combining the output channels with the mixing matrix $\mathbf{A}$ to generate two estimated signals. The oracle result is obtained by choosing the best match between the ground truth signals and the remixed signals among all mixing matrices. This oracle remixing achieves 0.8 dB improvement over the energy selection mechanism, highlighting the over-separation and selection problem of baseline MixIT. As shown in Table~\ref{tab:ts-mixit_anechoic}, our model outperforms even the MixIT model with oracle selection.
Since the final student model has a smaller number of output channels than the MixIT model, this result suggests that TS-MixIT effectively addresses the over-separation issue, therefore the target signals in the student model can more easily and accurately be selected.

When some mixture of the mixtures are created with single source signals, both the MixIT model and the proposed model achieve large gains over models trained with only two source mixtures.
This observation is consistent with that found in~\cite{wisdom2020unsupervised}, although the simulation strategies of using available single source samples are different. Notably, our method still provides a consistent gain over the MixIT model, and outperforms a system trained on a fraction (10\%) of fully supervised data.



\begin{table}[t]

\centering
\setlength{\tabcolsep}{4pt}
\caption{Results with model fine-tuning and knowledge distillation for anechoic single-channel separation}
\label{tab:finetune_distill_single}
\begin{tabu}{lcc}
\hline
\textbf{Method}      & \textbf{Architecture}      & \textbf{SI-SNRi (dB)}   \\ \hline
{TS-MixIT (Proposed)}           & Conv-TasNet                & 12.6         \\
{+Distill}                      & DPRNN                     & 12.9          \\
{+Fine\_tuning (10\%)}           & Conv-TasNet                & 13.2                    \\
{+Fine\_tuning (10\%)+Distill} & DPRNN                   & 14.3                  \\ \hline
{MixIT (Energy)} & DPRNN                  & 4.0                \\

\rowfont{\color{gray}}
{Supervised (10\%)}  & Conv-TasNet                & 11.0                    \\
\rowfont{\color{gray}}
{Supervised (10\%)}  & DPRNN                      & 11.7                 \\ 
\rowfont{\color{gray}}
{Supervised (100\%)}         & Conv-TasNet                & 15.3                    \\ \hline

\end{tabu}
\vspace*{-3mm}
\end{table}


The effect of model fine-tuning and knowledge distillation is assessed next (Table~\ref{tab:finetune_distill_single}).
To make full use of the small amount of available supervised data, the student model trained with 1or2-source is fine-tuned with 10\% of the official training data. Results show that the fine-tuned model outperforms both the student model and the supervised model trained with the same amount of data (line 6 in Table~\ref{tab:finetune_distill_single}). To further improve the network for the separation task, the knowledge learned from the fine-tuned model is distilled to the DPRNN model by using the whole unsupervised data again (line 4 in Table~\ref{tab:finetune_distill_single}). Since the DPRNN is a better designed model for the specific separation task, it can exploit the knowledge from the teacher more effectively and achieves a gain of 14.3~dB SI-SNR. We have also observed that directly training DPRNN with MixIT using the energy selection criterion leads to poor performance (line 5 in Table~\ref{tab:finetune_distill_single}), indicating that the teacher-student approach is important for successfully exploiting other model architectures.

\subsection{Multi-channel separation}

Table~\ref{tab:ts-mixit_task2} shows the performance of the multi-channel separation network in the noisy and reverberant condition. The results show that the spatial information helps improve the separation accuracy of models trained with the MixIT criterion.
Furthermore, TS-MixIT significantly outperforms the baseline for both data mixing approaches. Compared with the baseline MixIT, TS-MixIT alleviates the selection and over-separation problems by reducing the number of output channels from eight to two.


\begin{table}[!b]
\vspace*{-3mm}
\centering
\setlength{\tabcolsep}{4pt}
\caption{Results for 2-channel denoising and separation using unsupervised learning (Conv-TasNet architecture)}
\label{tab:ts-mixit_task2}
\begin{tabu}{lcccc}
\hline
\textbf{System}      & \textbf{Mixes} & \#nchs      & \textbf{M} & \textbf{SI-SNRi (dB)} \\ \hline
{MixIT (Energy)}     & 2-src          & 1           & 8          & 3.5                   \\
{MixIT (Energy)}     & 2-src          & 2           & 8          & 5.6                   \\
{TS-MixIT (Proposed)}           & 2-src          & 2           & 2          & 6.5                   \\ \hline
{MixIT (Energy)}     & 1or2-src       & 2           & 8          & 6.1                   \\
{TS-MixIT (Proposed)}           & 1or2-src       & 2           & 2          & 7.7                  \\ \hline
\rowfont{\color{gray}}
{Supervised (10\%)}  & -       & 2           & 2          & 8.5                  \\  \hline
\end{tabu}

\end{table}



\begin{table}[!t]

\centering
\setlength{\tabcolsep}{2pt}
\caption{Results with model fine-tuning and knowledge distillation for noisy multi-channel separation}
\label{tab:finetune_distill_multi}
\begin{tabu}{lcc}
\hline
\textbf{Method}      & \textbf{Architecture}      & \textbf{SI-SNRi (dB)} \\ \hline
{TS-MixIT (Proposed)}           & Conv-TasNet                & 7.7                  \\

{+Fine\_tuning (10\%)} & Conv-TasNet                & 9.2                  \\
{+Fine\_tuning (10\%)+Distill}          & U-ConvBlock                & 9.7                \\\hline
\rowfont{\color{gray}}
{Supervised (10\%)}  & Conv-TasNet                & 8.5                  \\
\rowfont{\color{gray}}
{Supervised (10\%)}  & U-ConvBlock                & 8.5               \\ 
\rowfont{\color{gray}}
{Supervised} (100\%)         & Conv-TasNet                & 11.1                  \\ \hline
\end{tabu}

\end{table}

Table~\ref{tab:finetune_distill_multi} reports the results of model fine-tuning and knowledge distillation in the multi-channel case. We use the noisy and reverberant mixtures as the input to the separation model and the reverberant sources as targets. This is a more challenging task since the model learns to do both denoising and speaker separation. With 10\% fine-tuning (line 2 in Table~\ref{tab:finetune_distill_multi}), the model achieves a large gain over the unsupervised student model. This can be explained as the unsupervised pre-trained model is encouraged to be more specific in the denoising and separation tasks with the supervised data. In addition, the fine-tuned model outperforms the supervised model trained with the same amount of paired data (line 4 in Table~\ref{tab:finetune_distill_multi}), indicating that the model can benefit from the unsupervised pre-training.
The previous fine-tuned model can also be distilled to an U-ConvBlock based model as shown on the third line of Table~\ref{tab:finetune_distill_multi}. This yields an additional 0.5 dB SI-SNR improvement.

Finally, the proposed system is evaluated in terms of ASR accuracy. The results are presented in Table~\ref{tab:whamr_wer}, and they show that the TS-MixIT model has significantly reduced the WER compared with the unprocessed mixtures for both matched and mismatched AMs.
Using the mismatched ASR system (AM1), there is no significant difference between the vanilla TS-MixIT and the one enhanced with fine-tuning and knowledge distillation, however, a large accuracy improvement is achieved for the latter algorithm by using the matched ASR model (AM2). The results with a fully supervised separation model is also provided for comparison purposes.
In the future, we plan to extend the semi-supervised approach to a joint denoising, dereverberation, and separation task and evaluate with data recorded in realistic environments such as, for example, the CHiME-5 corpus~\cite{barker2018fifth}.

\begin{table}[t]
\centering
\caption{Speech recognition results}
\label{tab:whamr_wer}
\begin{tabu}{lccc}
\hline
\multirow{2}{*}{\textbf{System}} & \multirow{2}{*}{\#nchs} & \multicolumn{2}{c}{\textbf{WER(\%)}} \\ \cline{3-4} 
                                 &                        & AM1               & AM2              \\ \hline
Mixture                          & -                      & 79.1              & 76.7             \\
{TS-MixIT (Proposed)}                       & 2                      & 42.4      & 35.7                \\
{+Fine\_tuning (10\%)+Distill}              & 2                      & 43.7      & 28.5        \\ \hline
\rowfont{\color{gray}}
{Supervised (100\%)}             & 2                      & 34.8      & 24.7        \\
\rowfont{\color{gray}}
Noisy Oracle                     & -                      & 19.8              & 19.8             \\ \hline
\end{tabu}
\vspace*{-4mm}

\end{table}

\section{Conclusions}
\label{sec:conclude}

This paper has presented a novel semi-supervised approach for end-to-end speech separation. By using the MixIT model as a teacher to train a student model with a smaller number of output channels, the method resolves the teacher's over-separation problem. With a limited amount of supervised data for fine-tuning and further improvement via model distillation, the separation performance of our semi-supervised method is comparable to that of a fully supervised system using ten times the amount of supervised data. 

\section{Acknowledgements}
The first author gratefully acknowledges financial support from Toshiba Europe Limited.



\bibliographystyle{IEEEtran}

\bibliography{ref}

\end{document}